\pdfoutput=1
\documentclass[twocolumn,english,superscriptaddress]{revtex4-1}

\usepackage{babel}
\usepackage{amsmath}
\usepackage{amssymb}
\usepackage{graphicx}

\newcommand{\appropto}{\mathrel{\vcenter{
  \offinterlineskip\halign{\hfil$##$\cr
    \propto\cr\noalign{\kern2pt}\sim\cr\noalign{\kern-2pt}}}}}

\begin{document}

\title{Kibble-Zurek exponent and chiral transition of the period-4 phase of Rydberg chains}

\author{Natalia Chepiga}
\affiliation{Institute for Theoretical Physics, University of Amsterdam, Science Park 904 Postbus 94485, 1090 GL Amsterdam, The Netherlands}
\author{Fr\'ed\'eric Mila}
\affiliation{Institute of Physics, Ecole Polytechnique F\'ed\'erale de Lausanne (EPFL), CH-1015 Lausanne, Switzerland}

\date{\today}


\begin{abstract}
Chains of Rydberg atoms have emerged as an amazing playground to study quantum physics in 1D. Playing with inter-atomic distances and laser detuning, one can in particular explore the commensurate-incommensurate transition out of charge-density waves through the Kibble-Zurek mechanism, and the possible presence of a chiral transition with dynamical exponent $z>1$. Here we address this problem theoretically with effective blockade models where the short-distance repulsions are replaced by a constraint of no double occupancy. For the period-4 phase, we show there is an Ashkin-Teller transition point with exponent $\nu=0.78$ surrounded by a direct chiral transition with a dynamical exponent $z=1.14$ and a Kibble-Zurek exponent $\mu=0.4$. For Rydberg atoms with a van der Waals potential, we suggest that the experimental value $\mu=0.25$ is due to a chiral transition with $z\simeq 1.9$ and $\nu\simeq 0.47$ surrounding an Ashkin-Teller transition close to the 4-state Potts universality. 
\end{abstract}
\pacs{
75.10.Jm,75.10.Pq,75.40.Mg
}

\maketitle


\section{Introduction}

Understanding the nature of quantum phase transitions in low-dimensional systems is one of the central topics in condensed matter physics\cite{giamarchi,tsvelik}.  Over the last decades, the combination of conformal field theory in 1+1D\cite{cardy,difrancesco} and advanced numerical techniques such as the density matrix renormalization group algorithm (DMRG)\cite{dmrg1,dmrg2,dmrg3,dmrg4} has proven to be extremely powerful in coming up with theoretical predictions for numerous fascinating critical phenomenas. In that respect, modern quantum simulators based on Rydberg atoms trapped with optical tweezers offer a remarkably rich experimental playground to further investigate quantum physics in 1D. 
In particular, in a recent experiment\cite{kibble_zureck}, the quantum critical dynamics of a chain of Rubidium atoms $^{87}$Rb with programmable interactions has been probed. The atoms are excited to a Rydberg state by a homogeneously applied laser with Rabi frequency $\Omega$, while the laser detuning $\Delta$ controls the population of excited atoms. The quantum many-body Hamiltonian of the system can be written in terms of hard-core bosons (i.e. bosons with no more than one particle per site) as:
\begin{eqnarray}
  H_{\text{Ryd}}&=&\sum_i \left[-\frac{\Omega}{2}(d^\dagger_i+d_i)-\Delta n_i+\sum_{R=1}^{+\infty} V_R n_in_{i+R}\right]
  \label{eq:hamilt}
\end{eqnarray}
where the van der Waals potential between Rydberg states decays as
\begin{eqnarray}
  V_R&=&V \left(\frac{1}{R}\right)^6.
  \label{eq:potential}
\end{eqnarray} 
The competition between the detuning term $\Delta$ that favors a high density of Rydberg states and the blockade leads to a sequence of lobes of charge density-wave phases with fixed periodicities. In general these periodicities can be any rational number, and  in the classical limit $\Omega\rightarrow 0$ they form a Devil's staircase\cite{Bak_1982}, but at finite values of $\Delta$, the phase diagram is dominated by lobes of integer periodicities $p=2,3,4,...$ \cite{kibble_zureck,rader2019floating,Bernien2017}, surrounded at least partially by a critical floating phase for $p\ge 3$\cite{rader2019floating}. However, according to recent experiments in which the detuning frequency has been swept for various interatomic distances $a$, this floating phase cannot be present along the whole boundary for $p=3$ and $4$ since a direct transition with a non-integer dynamical exponent $z$ larger than 1 has been detected in the vicinity of the tip of the lobe\cite{kibble_zureck}.

The transition out of a period-$p$ phase is an example of commensurate-incommensurate transition, a problem
with a long history that goes back to the investigation of adsorbed monolayers on surfaces\cite{Den_Nijs,birgeneau,SelkeExperiment}. In these systems the role of Rydberg atoms is played by domain walls between periodic phases, and quite remarkably the melting of these periodic phases is a very subtle problem that has not yet received a full solution. For $p=2$, the transition is known to be generically Ising, while for $p\ge 5$ it is a 2-step process through 
a Luttinger liquid phase (called a floating phase in the context of adsorbed monolayers) if it is not first order. The difficult cases are precisely $p=3$ and $p=4$. In these cases, the order-disorder transition along the commensurate line of the disordered phase is expected to be continuous in the 3-state Potts universality class for $p=3$, and in the Ashkin-Teller universality class (see below) for $p=4$. Away from this line, the disordered phase is incommensurate. As pointed out by Ostlund\cite{Ostlund} and Huse\cite{huse}, this introduces a chiral perturbation, and the open problem is to understand the effect of this chiral perturbation on the transition.

For $p=3$, the chiral perturbation is always relevant, and the question is whether it immediately opens a floating phase away from the Potts point, or whether the transition remains direct and continuous for a while, but in a new chiral universality class, as suggested by Huse and Fisher\cite{HuseFisher}, with a dynamical exponent $z>1$. Numerical\cite{Selke1982,Duxbury} and experimental evidence\cite{birgeneau,SelkeExperiment} in favor of this possibility has been obtained in the eighties and early nineties in the context of adsorbed layers, and very recently in the context of Rydberg atoms\cite{Bernien2017,kibble_zureck,samajdar,prl_chepiga,dalmonte}.

\begin{figure}[t!]
\centering 
\includegraphics[width=0.45\textwidth]{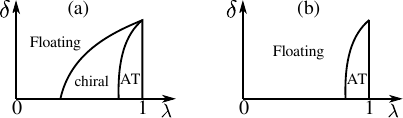}
\caption{Sketches of the possible nature of the transition out of the period-4 phase as a function of the Ashkin-Teller parameter $\lambda$ and of the chiral perturbation $\delta$ (a) with and (b) without a chiral transition. The width of the Ashkin-Teller (AT) phase has been exaggerated for visibility.}
\label{fig:theory}
\end{figure}

\textbf{The case of the period-4 phase}. For $p=4$, the situation is even richer because the chiral perturbation is not always relevant\cite{Den_Nijs}. With four degrees of freedom, there is in fact a family of universal classes described by the Ashkin-Teller model in which the local degrees of freedom are described by two Ising spins $\sigma_i\otimes \tau_i$ coupled by an interaction $\sigma_i\sigma_j+\tau_i\tau_j+\lambda \sigma_i\tau_i\sigma_j\tau_j$.
The asymmetry parameter $\lambda$ controls the relevance of the chiral perturbation. 
 
Indeed, according to Schulz\cite{schulz}, the cross-over exponent $\phi$ of the chiral perturbation for the Ashkin-Teller model is given by
\begin{equation}
\phi=\frac{3\nu}{2}+\frac{1}{4}-\frac{\nu^2}{2\nu-1}
\end{equation}
where $\nu$ is the exponent of the correlation length
The chiral perturbation is relevant if $\phi>0$, i.e. if $\nu>\nu_c=(1+\sqrt{3})/4=0.683...$, irrelevant otherwise.
Now, the exponent $\nu$ is known exactly as a function of $\lambda$\cite{kohmoto,obrien}:
\begin{equation}
\nu=\frac{1}{2-\frac{\pi}{2}[\arccos(-\lambda)]^{-1}}
\label{eq:nu_lambda}
\end{equation}
At $\lambda=0$, the model is known as the four-state clock model and corresponds to two decoupled transverse-field Ising chains. In that case, $\nu=1$: The chiral perturbation is relevant, and it is known to drive the system immediately into a critical phase. 
By contrast, at the 4-state Potts model ($\lambda=1$), $\nu=2/3<\nu_c$: The chiral perturbation is irrelevant. The critical value of $\lambda$ below which the chiral perturbation becomes relevant is given by
\begin{equation}
\lambda_c=-\cos \frac{\pi (\sqrt{3}+1)}{4(\sqrt{3}-1)}\simeq 0.9779...
\end{equation}
As long as the chiral perturbation is irrelevant, a line of continuous transition in the Ashkin-Teller universality class can be expected around the commensurate line until the chiral perturbation becomes relevant. Then the situation is similar to the $p=3$ case, with again the possibility of a chiral transition before a floating phase appears, as emphasized by Huse and Fisher\cite{HuseFisher1984}. These two possibilities are summarized in Fig.\ref{fig:theory}. 

Now, for $p=4$, the experimental results on Rydberg atoms\cite{Bernien2017,kibble_zureck} are compatible with a continuous transition, with a Kibble-Zurek exponent $\mu\simeq 0.25$. This exponent is related to $\nu$ by the relation $\mu=\nu/(1+\nu z)$, where $z$ is the dynamical exponent. Since between the clock model ($\lambda = 0$) and the Potts model ($\lambda = 1$) the exponent $\nu$ decreases from $1$ to $2/3$, the Kibble-Zurek exponent should be between $1/2$ and $2/5$ if the dynamical exponent was equal to 1. So, according to these experiments, the dynamical exponent has to be larger than 1. This implies that the transition should be a chiral Huse-Fisher transition, i.e. that the scenario of Fig.\ref{fig:theory}(a) is realized. 

In this paper, we investigate this problem in the context of an effective model for the period-4 phase, the quantum hard-boson model with two-site blockade (see below). We show that: (i) the transition along the commensurate line is sufficiently far from the 4-state Potts point to ensure that the chiral perturbation is relevant;  (ii)   there is an intermediate floating phase far enough from this point; (iii) there is evidence in favour of a small region of chiral transition in between for which we estimate the dynamical exponent and the Kibble-Zurek exponent. Implications for the original model of Eq.\ref{eq:hamilt} and for the experiments on Rydberg atoms are also discussed.

\textbf{The blockade models.} 
Because of the very steep increase of the van der Waals potential at short distance, the simultaneous excitation of atoms at a distance smaller than the so-called Rydberg radius $R_b\equiv (V_1/\Omega)^{1/6}$ is essentially excluded, a phenomenon known as Rydberg blockade. This means that, on a chain with lattice parameter $a$, the interaction between sites $i$ and $j$ can be considered to be infinite if $i-j\le r$, where $r$ is the largest integer satisfying $r<R_b/a$. Keeping only the dominant next-to-blockade repulsion leads to the following effective Hamiltonian:
\begin{equation}
  H=\sum_i-\frac{\Omega}{2}(d^\dagger_i+d_i)-\Delta n_i+V_{r+1} n_in_{i+r+1},
  \label{eq:blockade}
\end{equation}
where $d_i$ ($d^\dagger_i$) is an annihilation (creation) operator that acts in a constrained Hilbert space:
\begin{equation}
  n_i(n_i-1)=n_in_{i+1}=...=n_in_{i+r}=0.
  \label{eq:constraints}
\end{equation}
We will refer to this model as the $r$-site blockade model.
When $r=1$, it reduces to the original hard-boson model introduced by Fendley et al\cite{fendley}. Note also that a constrained Hilbert space equivalent to $r=2$ has been introduced by Huijse et al in the context of a supersymmetric model on a zig-zag ladder\cite{huijse}. 
Quite generally, the $r$-site blockade model allows one to discuss period $p=r+1$ and $p=r+2$ phases and their surrounding.
The main advantage of these constrained models is that their Hilbert space grows much more slowly than that of the original model of Eq.(\ref{eq:hamilt}), and simulations can be performed on systems large enough to keep track of small changes in the incommensurability and to identify the critical behaviour at the transition. 
Interestingly, the $r$-site blockade model can be seen as the limit of Rydberg atoms with infinitely fast decaying interactions. Indeed, 
if we consider the model  of Eq.(\ref{eq:hamilt}) with the interaction
\begin{eqnarray}
V_R&=&V_{r+1} \left(\frac{r+1}{R}\right)^\alpha,
  \label{eq:alpha}
\end{eqnarray}
the $r$-site blockade model corresponds to $\alpha\rightarrow +\infty$ while the $1/R^6$ model of Eqs.(\ref{eq:hamilt},\ref{eq:potential}) is recovered for $\alpha=6$. 

\section{Results}

\begin{figure}[t!]
\centering 
\includegraphics[width=0.45\textwidth]{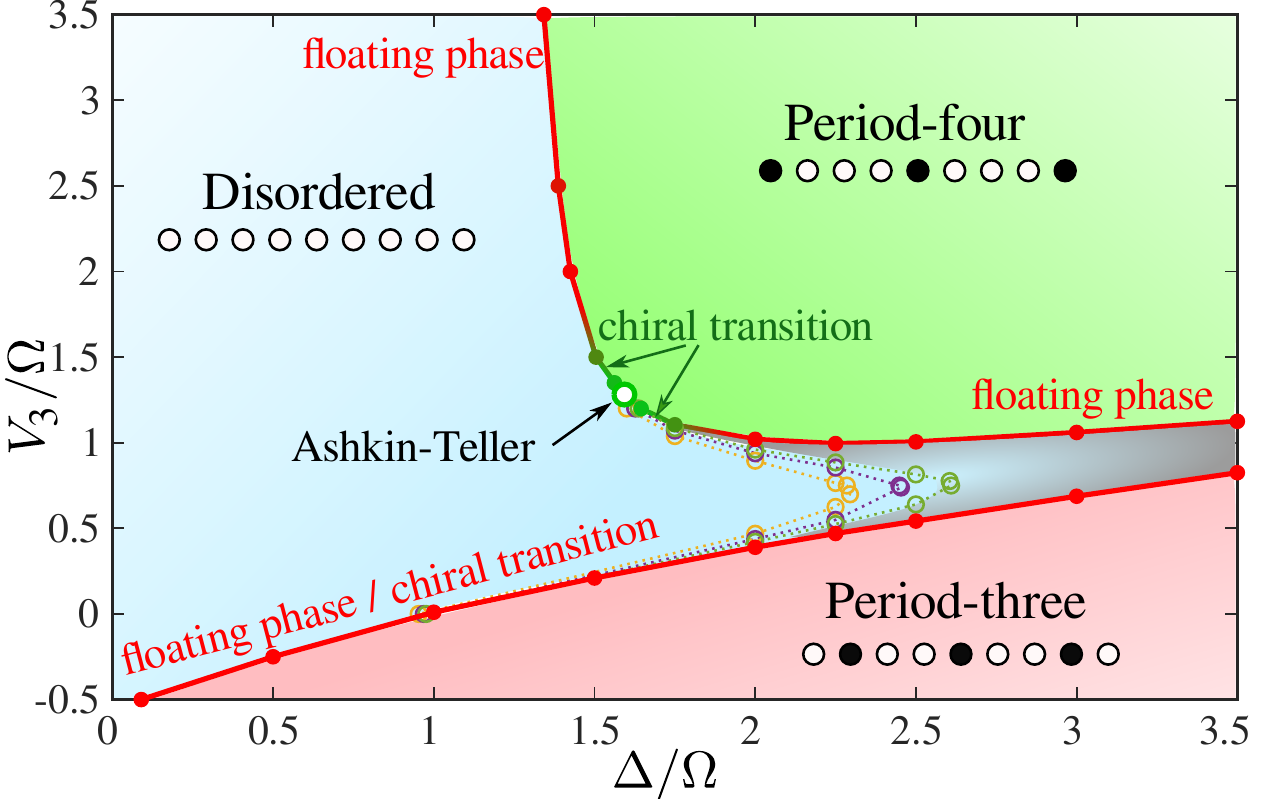}
\caption{Phase diagram of a hard-boson model with two-site blockade.
At the commensurate transition point located at $\Delta/\Omega\simeq1.593$ and $V_3/\Omega\simeq 1.2839$ the transition is in the Ashkin-Teller universality class with $\lambda\simeq0.57$ (open green circle). Away from it but not too far, our results are consistent with a chiral transition in the Huse-Fisher\cite{HuseFisher1984} universality class. Further away from the Ashkin-Teller point we detect an intermediate floating phase. 
The transition to the $p=3$ phase is through a direct chiral transition or through a the floating phase\cite{prl_chepiga}. 
For large $\Delta$ the two floating phases adjacent to the  $p=3$ and $p=4$ phases get close to each other and eventually merge into a single floating phase connecting the two ordered phases. Dotted lines state for $\xi=50$ (yellow), $100$ (purple) and $200$ (green).
}
\label{fig:phasediag}
\end{figure}

\textbf{Overview of the phase diagram for $p=4$.} As a first step towards the period-4 phase of Rydberg atoms, let us now turn to the properties of the 2-site blockade model. Our numerical results have been obtained with a state-of-the-art DMRG algorithm \cite{dmrg1,dmrg2,dmrg3,dmrg4} that explicitly implements the constraints (see Methods for details about the algorithm). They are summarized in the phase diagram of Fig.\ref{fig:phasediag}. There are three main phases: a disordered phase with incommensurate short-range correlations, and two ordered commensurate phases with period 3 and 4 respectively. 
Note that these three main phases  have been accessed in recent Rydberg atom experiments\cite{kibble_zureck,Bernien2017}.
 There are also small floating phases close to the ordered phases. In particular, for large values of $\Delta$, there are two floating phases at the boundaries of the period-three and period-four phases that come closer and create an area of extremely high correlation length. It is therefore probable that the disordered phase eventually disappears and that, for some parameter range, the two ordered phases are connected through a single floating phase, as suggested in Ref.\cite{rader2019floating,Bernien2017} for the model of Eq.(\ref{eq:hamilt}). Due to the exponential growth of the correlation length at the Kosterlitz-Thouless\cite{Kosterlitz_Thouless} phase transition, an accurate investigation of this scenario would require simulations beyond our current limitations. 

\begin{figure}[t!]
\centering 
\includegraphics[width=0.42\textwidth]{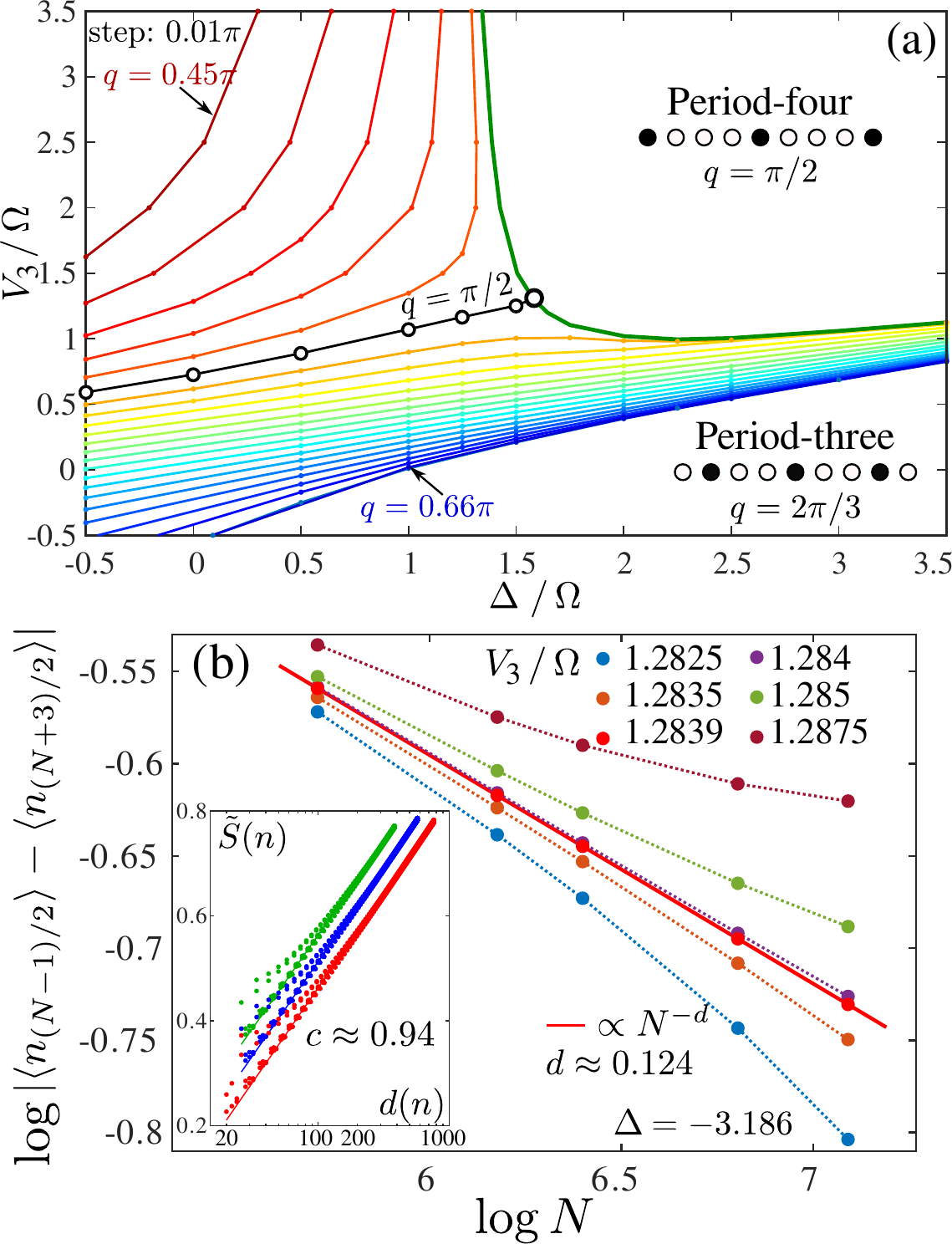}
\caption{ Identification of the conformal point. (a) Phase diagram with equal-$q$ lines in the disordered phase extracted for $N=601$ (systematically) and for $N=1201$ (eventually, in the vicinity of the critical lines). The location of the Ashkin-Teller point has been determined as the crossing point of the critical (green) line and the $q=\pi/2$ line (black, open circles). (b) Finite-size scaling of the amplitude of the oscillations in on-site boson density in the middle of the finite-size chain. The separatrix corresponds to the critical point.
Inset: Scaling of the entanglement entropy with the conformal distance $d(n)$ after removing the Friedel oscillations,
leading to a central charge $c\simeq0.94$.}
\label{fig:location}
\end{figure}

\textbf{Commensurate line.} The transition out of the period-four phase is the main focus of the rest of this section. Our first task is to locate the point on the phase boundary where the chiral perturbation vanishes, hence where the transition can be expected to be described by a conformal field theory. 
Note that for the original hard-boson model of Fendley et al \cite{fendley} this was not necessary because the 3-state Potts belongs to an integrable line, and its location is known exactly.
Here this is not the case, but we can expect this point to be located at the intersection of the phase boundary and of the line with wave vector $q=\pi/2$ since along this line the correlations remain commensurate in the disordered phase so that there is not chiral perturbation. To achieve this goal, we have determined the lines of constant incommensurate wave-vector $q$. They are depicted in Fig.\ref{fig:location}(a).The line $q=\pi/2$ enters the period-four phase at $\Delta/\Omega\simeq 1.593$. An accurate estimate of the second coordinate has been obtained by a finite-size scaling of the order parameter. It turns out that open boundary conditions favor a boson on the first and last sites. This effectively acts as a conformally-invariant fixed boundary condition at the critical point and induces Friedel oscillations in the local boson density. According to boundary conformal field theory the profile of these oscillations on a finite-size chain is given by $\propto [N\sin(\pi j/N)]^{-d}$, where the scaling dimension $d=1/8$ for the Ashkin-Teller model\cite{PhysRevB.91.165129}. By scanning $V_3/\Omega$ for $\Delta/\Omega=1.593$, we identify a separatrix in the log-log scaling at $V_3/\Omega=1.2839$ as shown in Fig.\ref{fig:location}(b). The slope corresponds to $d\simeq0.124$, in excellent agreement with the scaling dimension $d=1/8$. As a further check that this is a critical point, we have extracted the central charge by fitting the profile of the reduced entanglement entropy to the Calabrese-Cardy formula (see Methods), leading to a central charge $c\simeq 0.94$, within $6\%$ of the conformal field theory prediction  $c=1$.

 \begin{figure}[t!]
\centering 
\includegraphics[width=0.4\textwidth]{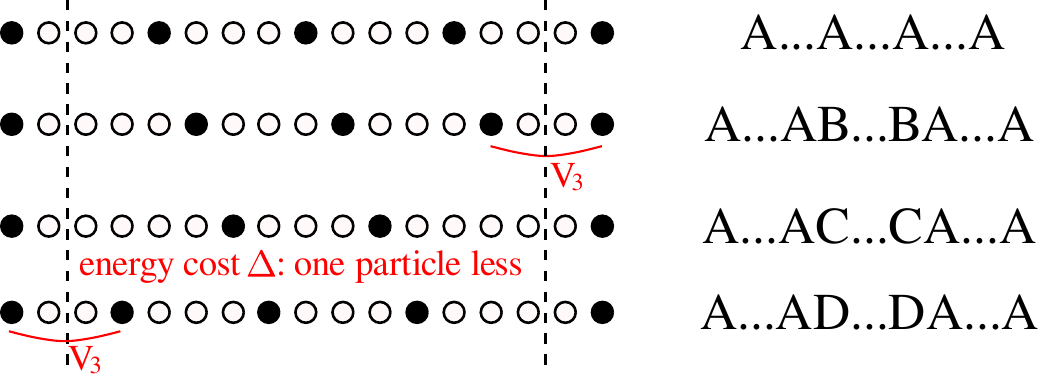}
\caption{Asymmetry of domain walls in the model with two-site blockade.  For $p=4$, domains with B or D inside A cost an energy $V_3$ while domains with C inside cost an energy $\Delta>V_3$ since there is one particle less, leading to an asymmetry in the effective transverse field term.  }
\label{fig:asymmetry}
\end{figure}

The correlation length of the hard-boson model can be simply obtained by fitting correlations, a straightforward task along the commensurate line (see Methods). The resulting correlation diverges at the critical point with an exponent $\nu\simeq 0.78$. This is the first indication that $\lambda$ must be significantly smaller than 1. This is actually quite natural. Indeed,
when $\lambda=1$, the model corresponds to the four-states Potts model with the same amplitude for all flipping processes, while for $\lambda<1$ two processes are favoured over the third one by the transverse field term. Such an asymmetry naturally appears in the hard-boson model due to the two-site blockade. From Fig.\ref{fig:asymmetry} one can see that domains B and D shifted by one site with respect to the bulk A cost less energy than the domain C shifted by two sites. 

One can also estimate $\lambda$ directly by comparing the excitation spectrum of the two-site blockade model with that of the quantum 1D version of the Ashkin-Teller model (see Methods). This leads to $\lambda \simeq 0.57$, in excellent agreement with $\nu=0.78$.
At that point, the chiral transition is relevant, with a crossover exponent $\phi\simeq 0.33$. This means in particular that, away from that point, the transition cannot be a standard continuous transition in the Ashkin-Teller universality class. Either a floating phase opens, or the transition becomes chiral.

\textbf{Chiral transition versus floating phase}. Quite generally, the incommensurate wave-vector $q$ is expected to approach the commensurate value $\pi/2$ with a critical exponent called $\bar{\beta}$. To the best of our knowledge the exact value of this critical exponent is not known for the Ashkin-Teller model, but Huse and Fisher\cite{HuseFisher1984} argue that $\bar{\beta}>\nu$. 
This implies that the product $\xi\times|\pi/2-q|$ decays to zero upon approaching the Ashkin-Teller transition. By contrast, if the transition is chiral, the equality $\bar{\beta}=\nu$ should hold, and $\xi\times|\pi/2-q|$ is expected to go to some finite value\cite{HuseFisher1984}. 
When the transition is Ashkin-Teller or chiral, the exponents of the correlation length $\nu$ in the disordered phase and $\nu'$ in the ordered phase should satisfy $\nu=\nu'$. By contrast, in the presence of an intermediate floating phase, the correlation length in the disordered phase diverges exponentially at a Kosterlitz-Thouless\cite{Kosterlitz_Thouless} transition, while the wave-vector $q$ remains incommensurate, so that the product $\xi\times|\pi/2-q|$ diverges. The commensurate-incommensurate transition between the floating and the ordered phases is then expected to be in the Pokrovsky-Talapov \cite{Pokrovsky_Talapov} universality class with critical exponent $\bar{\beta}=\nu^\prime=1/2$.

\begin{figure}[t!]
\includegraphics[width=0.45\textwidth]{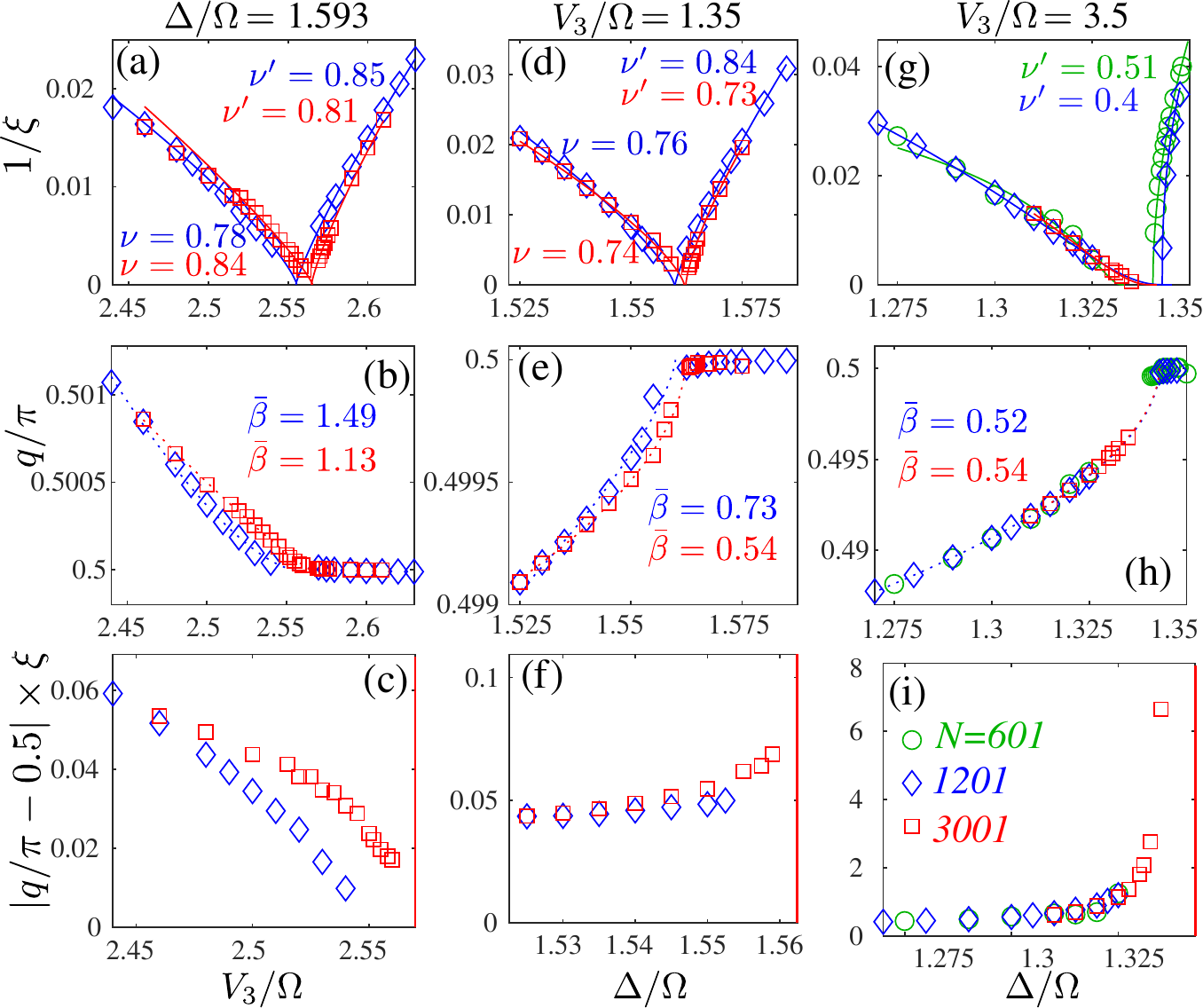}
\caption{Inverse correlation length $1/\xi$, wave-vector $q/\pi$ and product $\xi\times|\pi/2-q|$ at three different cuts across the transition. (a-c) Vertical cut through the Ashkin-Teller point at $\Delta/\Omega=1.593$; (d-f), (g-i) -Horizontal cuts at $V_3/\Omega=1.35$ and $V_3/\Omega=3.5$ respectively.  
Inside the $p=4$ phase, the correlation length is fitted with a power-law with critical exponent $\nu^\prime$. In the disordered phase, the correlation length is fitted either with a power-law with critical exponent $\nu$ (a,d), or with the KT form $\xi\propto \exp(\text{C}/\sqrt{g_{KT}-g})$ (g), where $g$ is the coordinate along the cut. The wave-vector $q$ is fitted with a power law with exponent $\bar{\beta}$ (dotted lines).
In the lower panels, the red lines indicate the boundary of the ordered phase.}
\label{fig:XiQ}
\end{figure}

In Fig.\ref{fig:XiQ} we take a closer look at three cuts across the transition. Let us start with the vertical cut  through the Ashkin-Teller point identified above at $\Delta/\Omega=1.593$. The critical exponents $\nu$ and $\nu^\prime$ are in good agreement with each other, and they are also in reasonable agreement with the value obtained for $\nu$ along the commensurate line and with the value of $\lambda$.
An accurate estimate of $\bar{\beta}$ is very difficult due to the proximity of the commensurate value of $q$ in the disordered phase. Nevertheless it is clear qualitatively, just looking at the curvature, that $\bar\beta$ is significantly larger than $\nu$, in agreement with Huse and Fisher\cite{HuseFisher1984}. As a consequence,  the product  $\xi\times|\pi/2-q|$ goes to zero at the critical point as shown in Fig.\ref{fig:XiQ}(c).

The next cut at $V_3/\Omega=1.35$, slightly away but very close to this Ashkin-Teller point, is presented in Fig.\ref{fig:XiQ}(d)-(f). The correlation length diverges as a power law with similar exponents on both sides of the transition, but, by contrast to the Ashkin-Teller point, the critical exponent $\bar{\beta}$ is much smaller than 1, a clear indication that the chiral perturbation changes the physics immediately away from the Ashkin-Teller point. Its value is comparable to $\nu$ and $\nu^\prime$, and accordingly, even if it increases slightly towards the transition, the product $\xi\times|\pi/2-q|$ seems to remain finite. The absence of divergence of the product $\xi\times|\pi/2-q|$ is a clear indication in favor of the Huse-Fisher universality class. However, as in the case $p=3$, an extremely narrow floating phase cannot be excluded. 

Further away from the commensurate point, the inverse of the correlation length decays in a very asymmetric way, as we show for the  horizontal cut at $V_3/\Omega=3.5$ in Fig.\ref{fig:XiQ}(g)-(i). The numerically extracted critical exponent $\nu^\prime$ is in reasonable agreement with the Pokrovsky-Talapov value $1/2$, while the  product $\xi\times|\pi/2-q|$ clearly diverges towards the transition.  The physics is very similar on the other side of the Ashkin-Teller line (see Supplementary Materials for more data).

\begin{figure}[t!]
\centering 
\includegraphics[width=0.48\textwidth]{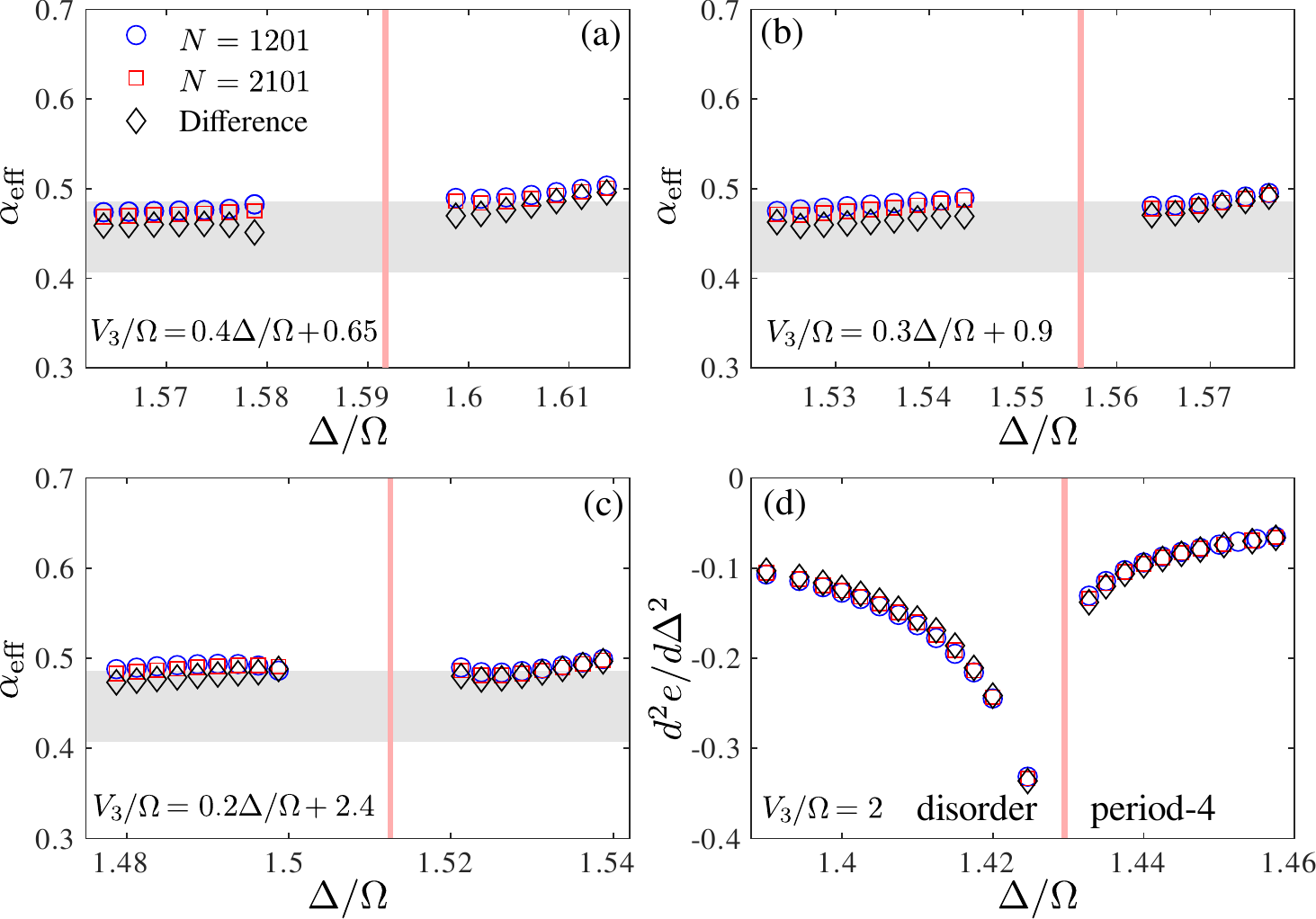}
\caption{ Effective critical exponent $\alpha$ across  (a) the Ashkin-Teller point and (b),(c) the chiral transition. (d) Second derivative of the energy per site with respect to $\Delta/\Omega$ for $V_3/\Omega=2$ around the Pokrovsky-Talapov transition. The results are extracted from the ground-state energy of a chain with $N=1201$ sites (blue circles) and $N=2101$ sites (red squares), and from the difference between the two (black diamonds). The grey area indicates the expected value of $\alpha$ for a critical exponent $\nu\approx0.78\pm0.02$.}
\label{fig:alpha}
\end{figure}

As a further check, we have investigated the behaviour of the second derivate of the ground state energy, the equivalent of the specific heat for quantum systems. If the transition is continuous, it is expected to diverge with the same exponent $\alpha$ on both sides of the transition, while if there is an intermediate floating phase it is expected to diverge with exponent $1/2$ at the Pokrovsky-Talapov transition when coming from the incommensurate phase, and to saturate with a logarithmic singularity on the other side.  As can be seen in Fig.\ref{fig:alpha}, the results are fully consistent with a single transition at and close to the commensurate line, and with an asymmetric behaviour far enough from it. According to hyperscaling, $\alpha$ should be related to $\nu$ at the Ashkin-Teller point by $\alpha=2(1-\nu)\simeq 0.44$, in good agreement with the numerical results of Fig. Fig.\ref{fig:alpha}(a). Interestingly, $\alpha$ barely changes as long as the transition is continuous, a fact already noticed and rigorously established for integrable and self-dual versions of the 3-state chiral Potts model\cite{albertini,Baxter1989,Cardy1993}.

\textbf{Kibble-Zurek mechanism and dynamical exponent}.  To estimate the Kibble-Zurek exponent  $\mu=\nu/(1+\nu z)$, we need both the dynamical exponent $z$ and the correlation length exponent $\nu$. Along the commensurate line, the transition is in the Ashkin-Teller universality class and has conformal invariance, so $z=1$. The estimate $\nu=0.78$ then leads to a Kibble-Zurek exponent $\mu=0.44$. Away from the Ashkin-Teller point, the dynamical exponent $z$ can be extracted from $\nu$ and $\alpha$ according to the hyperscaling relation for anisotropic systems: $\nu+z\nu=2-\alpha$. For the cut of Fig.\ref{fig:XiQ}(d), $\nu=0.73$. Assuming that $\alpha$ keeps the value $\alpha=0.44$ of the Ashkin-Teller point, in agreement with the discussion above, leads to $z=1.14$. Across this cut, the Kibble-Zurek exponent is thus given by $\mu=0.4$, significantly smaller than across the Ashkin-Teller transition. This conclusion remains true even if we assume that $\alpha$ slightly increases away from the Ashkin-Teller point, as suggested by Fig.\ref{fig:alpha}.

\section{Discussion}

\label{sec:back_to_rydberg}

To make contact with experiments, let us briefly discuss the implications of the present results for the model with $1/R^6$ long-range interactions. 
As explained above, both models belong to the same family of models defined by Eq.(\ref{eq:alpha}). Let us estimate the effect of reducing $\alpha$ from $+\infty$ to $6$ for $r=2$ in the vicinity of the Ashkin-Teller transition. The critical values of the two-site blockade are given by $\Delta/\Omega=1.593$ and $V_3/\Omega=1.2839$. This value of $V_3$ corresponds to a Rydberg radius $R_b/a=3 (V_3/\Omega)^{1/6}=3.1276$, near the tip of the $p=4$ lobe where the experiments have been carried out\cite{kibble_zureck}, and where there is no evidence of a floating phase\cite{rader2019floating}. The critical value of $\Delta/\Omega=1.593$ is different from that of the Rydberg model at the tip of this lobe (around 2.39), but this is not surprising since $\Delta$ is the chemical potential in the bosonic language, and its critical value must be strongly affected by the details of the interactions. 

Now let us turn to the nature of the transition, assuming that there is a portion of boundary without floating phase. 
The physical reason behind $\lambda<1$ is the difference in energy cost of domains shifted by one or two sites with respect to the bulk (see Fig. \ref{fig:asymmetry}). In the very simple classical approximation $\delta E_{B,D}\simeq V_3 $ while $\delta E_{C}\simeq \Delta$. At the Ashkin-Teller critical point, these expressions lead to $\delta E_{B,D}\simeq 1.2839$ and $\delta E_{C}\simeq 1.593$ for the blockade model.
Taking into account longer-range interactions, the energy of domain walls for the Rydberg model can be estimated as $\delta E_{B,D}\simeq V_3-2V_4+V_5$ and $\delta E_{C}\simeq \Delta-3V_4+2V_7$. Assuming $V_3=V\simeq 1.2839$ and a $1/R^6$ decay, we get $\delta E_{B,D}\simeq 0.885$ and $\delta E_{C}\simeq 0.925$. The asymmetry is still present, but it is smaller, implying that the point where the chiral perturbation vanishes gets closer to the four-state Potts point. Therefore, there are two possibilities: (i) $\lambda$ is still smaller than $\lambda_c=0.9779$. Then the chiral perturbation remains relevant, and the transition immediately becomes chiral until a floating phase emerges; (ii)  the long-range interactions bring the Ashkin-Teller point close enough to the  four-state Potts point so that the chiral perturbation is irrelevant; then there will be an extended region of direct Ashkin-Teller transition, followed on both sides by a chiral transition, and ultimately by a floating phase. Since $\lambda_c\simeq 0.9779$ is very close to $1$, the first possibility (i) is more likely. More importantly, the fact that the asymmetry can be expected to be {\it reduced} by long-range interactions and not increased implies that, if anything, the Rydberg model is further away from the clock limit $\lambda=0$ where there would be an intermediate floating phase all along the boundary. So our conclusion that there is a portion of the boundary to the period-4 phase where the transition is direct and continuous in the chiral universality class before a floating phase opens can be considered as a prediction for the Rydberg model with $1/R^6$ interactions.

Note that the finite-size effects associated with the restricted number of Rydberg atoms in experiments\cite{kibble_zureck,Bernien2017} will, if anything, enlarge the portion without floating phase. Indeed, if, coming from the disordered phase, the floating phase starts at an incommensurate wave-vector $q$, its detection requires the size of the chain to be significantly larger than the period necessary to form at least one helix $N>2\pi/(q-\pi/2)$. So, for a finite-size system, the floating phase can only be detected further away from the commensurate line than in the thermodynamic limit, and the transition will look continuous in a larger parameter range, making the observation of this direct transition easier.

Finally, let us discuss briefly the consequences for the Kibble-Zurek experiment. If the transition is chiral, the scaling becomes anisotropic, but
if hyperscaling applies, the correlation exponent along the chains $\nu$ and the dynamical exponent $z$ are related by $\nu(1+z)=2-\alpha$. Let us further assume that, as for the two-site blockade model and the self-dual 3-state chiral Potts model, the specific heat exponent keeps the value it has at the  Ashkin-Teller critical point $\alpha=2(1-\nu_{\lambda})$. Then we get $\nu(1+z)=2\nu_{\lambda}$, where the Ashkin-Teller critical exponent $\nu_{\lambda}$is given by Eq.\ref{eq:nu_lambda}. This implies that $\nu$ and $z$ can be deduced from the Kibble-Zurek exponent $\mu$ and the asymmetry parameter $\lambda$ according to  $\nu=\left[\mu(1+2\nu_{\lambda})\right]/(1+\mu)$ and $z=(2\nu_{\lambda}-\mu)/\left[\mu(1+2\nu_{\lambda})\right]$. Taking the experimental value $\mu\simeq 0.25$ and assuming that $\lambda$ is close to 1, as suggested by the small asymmetry of domain walls for Rydberg atoms,
we get $z\simeq 1.9$ and $\nu \simeq 0.47$.
It will be interesting to see if these values can be confirmed by a direct numerical investigation of the model of Eqs.(\ref{eq:hamilt}-\ref{eq:potential}).

\section{Methods}

\begin{figure}[t!]
\centering 
\includegraphics[width=0.5\textwidth]{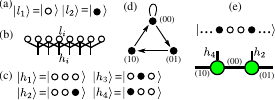}
\caption{(a) Local Hilbert space of the original model $|l_i\rangle$. The open (filled) circle stands for an empty (occupied) site. (b) Rigorous mapping onto a model with a local Hilbert space spanned over three consecutive hard bosons that consist of four states sketched in (c). The index of the new site corresponds to the index of the middle site. (d) Fusion graph for the recursive construction of the left environment; for the right environment the direction of the arrows should be inverted. (e) Example of the label assignment in MPS representation on two consecutive tensors (green circles) written for the selected state.}
\label{fig:fusion_graph}
\end{figure}

\textbf{Details about the algorithm.}
The size of the Hilbert space for a model with two-site Rydberg blockade can be calculated using a recursive relation $\mathcal{H}(N)=\mathcal{H}(N-1)+\mathcal{H}(N-3)$, with the first three elements of the sequence $\mathcal{H}(1)=2$, $\mathcal{H}(2)=3$ and $\mathcal{H}(3)=4$. 
So the growth of the Hilbert space with the system size $\mathcal{H}(N)\propto 1.466^N$ is much slower than $\mathcal{H}(N)\propto 2^N$ for an unconstrained model. 
In order to fully profit from the restricted Hilbert space we implement the blockade  explicitly into the DMRG. 
Recently it has been shown that the hard-boson model with $r=1$ can be rigorously mapped onto a quantum dimer model on a two-leg ladder\cite{scipost_chepiga} that provides a simple and intuitive way to encode the constraint into DMRG. Although this mapping is not valid for $r>1$, we can rely on the idea of auxiliary quantum numbers that would preserve the block-diagonal structure of the local tensors. This is achieved by a {\it rigorous} mapping onto an effective model that spans the local Hilbert space over three consecutive sites on the original lattice as shown in Fig.\ref{fig:fusion_graph}(b). The new local Hilbert space contains four states listed in Fig.\ref{fig:fusion_graph}(c). Because of the overlap, the three possible states of two shared sites can be used as a quantum label for the auxiliary bond between two consecutive sites of the new model. By adding a site, for example by increasing the left environment, one can change the quantum labels according to the fusion graph shown in Fig.\ref{fig:fusion_graph}(d). The fusion graph for the right environment can be obtained by inverting the arrows. An example of the label assignment is provided in Fig.\ref{fig:fusion_graph}(e).

At the next step, one has to rewrite the hard-boson model given by Eq.1 of the main text in terms of new local variables $|h_i\rangle$. For example, the boson occupation number operator $n_i$, which is also equal to $(1-n_{i-1})n_i(1-n_{i+1})$, can be written in the new local Hilbert space as a $4\times4$ matrix $\tilde{n}_i$ with the only non-zero element $\tilde{n}_i(3,3)=1$. The term $V_3n_{i-1}n_{i+2}$ can be written in the new Hilbert space  as a nearest-neighbor interaction $V_3\tilde{p}_i\tilde{q}_{i+1}$  where the only non-zero matrix elements of the operators $\tilde{p}$ and $\tilde{q}$ are given by $\tilde{p}(4,4)=1$ and $\tilde{q}(2,2)=1$. Finally the constrained flip term 
$$-\frac{\Omega}{2} (1-n_{i-2})(1-n_{i-1})(d_i^\dagger+d_i)(1-n_{i+1})(1-n_{i+2})$$
can be rewritten as a three-site operator
$$-\frac{\Omega}{2}(\tilde{a}_{i-1}\tilde{b}_i \tilde{c}_{i+1}+\mathrm{h.c}),$$ 
where the only non-zero matrix elements of the operators $\tilde{a}$, $\tilde{b}$, and $\tilde{c}$ are given by $\tilde{a}(1,2)=1$, $\tilde{b}(1,3)=1$, and $\tilde{c}(1,4)=1$.

With these definitions, the matrix product operator in the bulk takes the following simple form:
\begin{equation}
\left( \begin{array}{ccccccc}
\tilde{I} & . & . & . & . & . & .\\ 
\tilde{q} & . & . & . & . & . & .\\
\tilde{c} & . & . & . & . & . & .\\
\tilde{c}^\dagger & . & . & . & . & . & .\\
. & . & \tilde{b} & . & . & . & .\\
. & . & . & \tilde{b}^\dagger  & . & . & .\\
-\Delta \tilde{n} & V_3 \tilde{p} & . & .  & -\frac{\Omega}{2}\tilde{a} & -\frac{\Omega}{2}\tilde{a}^\dagger & \tilde{I}\\
  \end{array} \right),
\end{equation}
where dots mark zero entries of the tensor. Close to the edges one has to carefully modify the MPO to properly  encode the boundary terms. This requires the definition of local operators slightly different from those used in the bulk. 

There is yet another crucial point that we want to mention. The labels that we have introduced split the Hilbert space into blocks or sectors and therefore correspond to some conserved quantity.  For the hard-boson model with a single-site blockade, the quantum labels correspond to the parity of the domain walls. In the present case, the physical meaning of the conserved quantity is  not as obvious. However, the only relevant information for us 
is that the conservation of this abstract quantity requires at least three sites. In other words, by acting with any term (read flip term) on a two-site MPS, one necessary changes one of the out-going labels, while the flip term applied on three consecutive MPS keeps all external labels fixed. As a consequence, neither single- nor two-site DMRG routines are compatible with the presented constraint implementation, and one has to go for at least three-site updates. At a glance this might look costly with a local Hilbert space of dimension $4$ since it leads in principle to an MPO operator of size $7\times7\times 64\times64$. However, taking into account all the constraints on three sites, the projected three-site MPO is only of size $7\times7\times 9\times9$.

The explicit implementation of two-sites blockade allows us to reach systems with up to $N=3001$ sites systematically (and $N=4801$ sites occasionally), keeping up to 2000 states.

\textbf{Calabrese-Cardy formula.} 
According to Calabrese and Cardy\cite{CalabreseCardy} the entanglement entropy in finite-size chain with open boundary conditions scales with the block size $l$ as:
\begin{equation}
{S}_L(l)=\frac{c}{6}\ln d(l)+s_1+\log g,
\end{equation}
where $d(l)=\frac{2L}{\pi}\sin\left(\frac{\pi l}{L}\right)$ is the conformal distance; $s_1$ and $\log g$ are non-universal constants. 
The presence of Friedel oscillations caused by the fixed boundary conditions is also reflected in the entanglement entropy profile. 
In order to remove the oscillations we follow Ref.\cite{capponi} and construct the reduced entanglement entropy:
\begin{equation}
\tilde{S}_N(l)={S}_N(l)-\zeta\langle {\bf n}_{l-1}{\bf n}_{l+2}\rangle,
\end{equation}
where $\zeta$ is a non-universal constant in front of the leading local correlations between nearest allowed neighbors adjusted to best remove the oscillations.

\begin{figure}[t!]
\centering 
\includegraphics[width=0.45\textwidth]{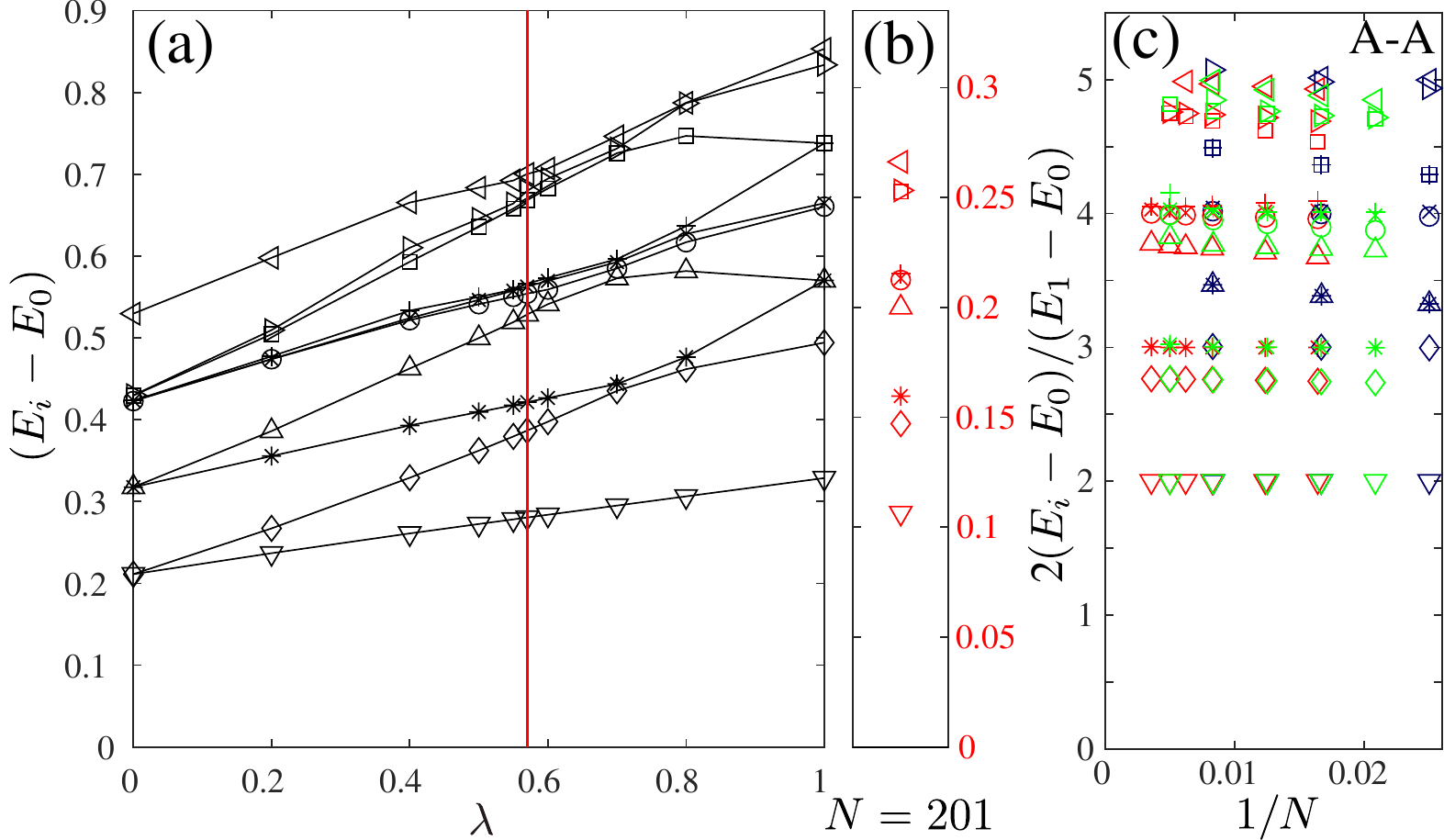}
\caption{ (a) Excitation spectrum of the Ashkin-Teller model as a function of $\lambda$ for  $N=60$ and $A-A$ boundary conditions. (b) Excitation spectrum of the hard-boson model with $N=201$ sites. (c) Conformal towers of states for hard-boson (red), Ashkin-Teller at $\lambda=0.57$ (green) and four-state Potts as Ashkin-Teller with $\lambda=1$ (blue). The tower is plotted with respect to the lowest excitation energy}
\label{fig:towers}
\end{figure}

\textbf{Comparison with the Ashkin-Teller model and estimate of $\lambda$.}
To estimate $\lambda$ directly, one can compare the excitation spectrum of the two-site blockade model with that of the quantum 1D version
of the Ashkin-Teller model defined in terms of Pauli matrices $\sigma^{x,z}$ and $\tau^{x,z}$ by the Hamiltonian:
\begin{multline}
H_{AT}=-\sum_{j=1}^N \left(\sigma_j^x+\tau_j^x+\lambda \sigma_j^x\tau_j^x\right)\\
-\beta\sum_{j=1}^{N-1} \left(\sigma_j^z\sigma_{j+1}^z+\tau_j^z\tau_{j+1}^z+\lambda \sigma_j^z\tau_j^z\sigma_{j+1}^z\tau_{j+1}^z\right),
\end{multline}
at its critical point $\beta=1$. The spectra have been obtained by targeting several states (up to 11) at every DMRG iteration\cite{dmrg_chepiga}. In Fig.\ref{fig:towers}(a) we show the energy spectrum of the Ashkin-Teller model for $N=60$ with fixed A-A boundary conditions.  We compare these results with the spectrum of the constrained model with $N=201$ sites. Since the velocity is a non-universal constant, one cannot compare the absolute values of the gap. However we find that the structure of the spectrum in the hard-boson model corresponds to the structure of the Ashkin-Teller spectrum at $\lambda\simeq 0.57$ (red line) (see Supplementary Materials for details). In Fig.\ref{fig:towers}(c) we further compare the finite-size scaling for hard boson (red) and Ashkin-Teller model at $\lambda=0.57$ (green) and at $\lambda=1$ (4-state Potts, blue) and the agreement with $\lambda=0.57$ is quite good.

The comparison can be made even more systematic by re-scaling both spectra with respect to the lowest excitation energy as explained in the supplemental materials.

We compute the energy spectrum in a chain with open and fixed boundary conditions. There are two reasons for that. First, DMRG is well known to be more efficient for open boundary conditions than for periodic ones. Second, the number of conformal towers of states that appears in the spectra of periodic or anti-periodic chains are usually larger than the number of towers selected by fixed boundary conditions.  However, we have to establish the correspondence between the different boundary conditions in the hard-boson model and in the original Ashkin-Teller model. 
In the hard-boson model, the simplest way to fix the boundary is to force the first and the last sites to be occupied. In the $p=4$ phase every fourth site is occupied by a boson. So each of the ground states, let us call them A, B, C and D, corresponds to the location of the occupied site $\mathrm{mod}\, 4$.   If the total number of sites is $4k+1$ the same state is favored at each edge, corresponding to the $A-A$ boundary condition in the Ashkin-Teller model. If the total number of sites is $4k$ or $4k+2$, we expect A-B and A-D boundary conditions. They are expected to give the same spectrum (assuming that states B and D have equal weight in the transverse field applied on A, while C has a factor $\lambda$). Finally, if the total number of sites is $4k+3$, we expect to observe the spectrum of the $A-C$ boundary condition. Numerical results for A-C and A-B/A-D boundary conditions are provided in the Supplemental materials.

\begin{figure}[t!]
\centering 
\includegraphics[width=0.4\textwidth]{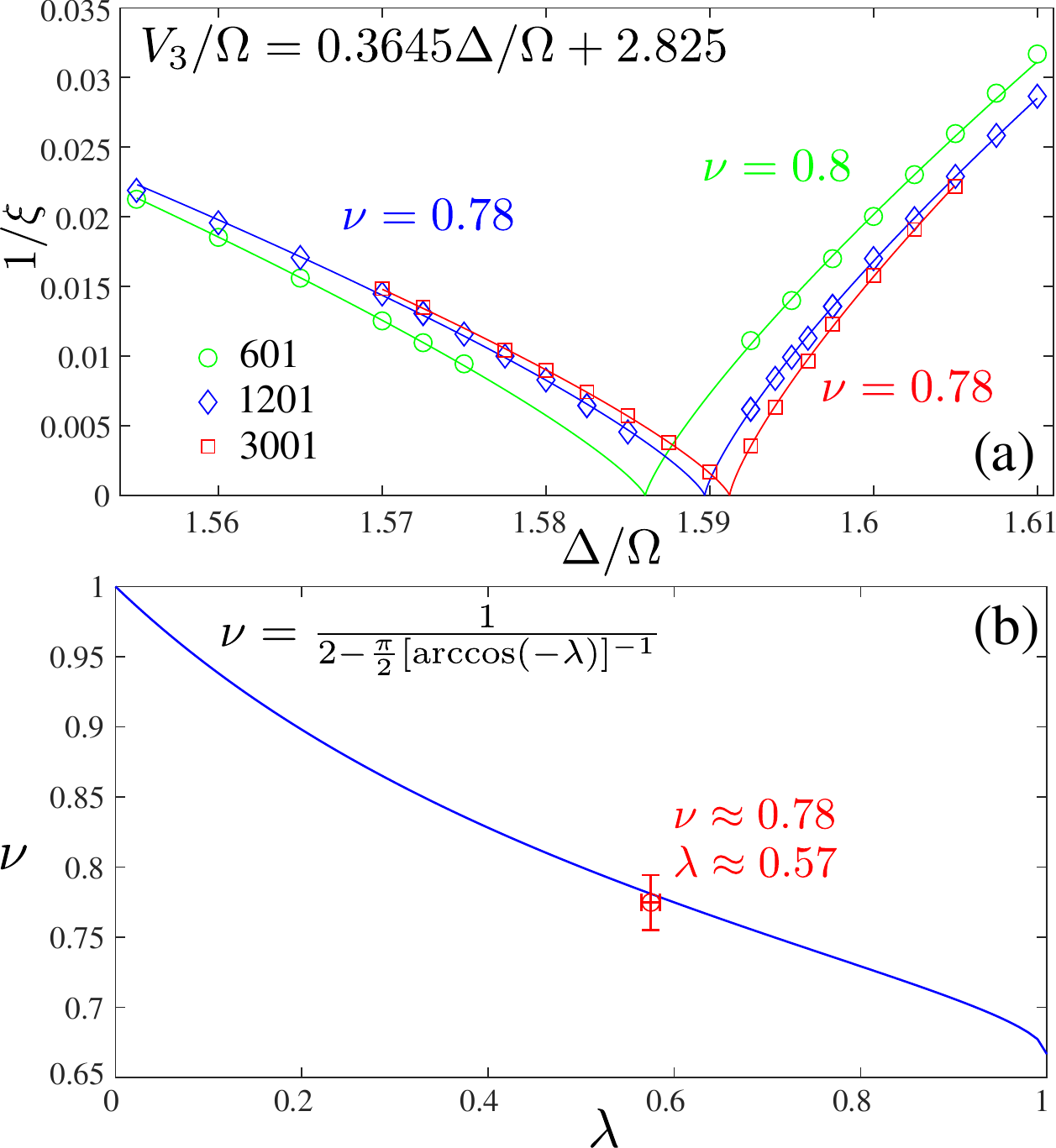}
\caption{ Comparison between the numerically extracted value of the critical exponent $\nu$ and the one of the Ashkin-Teller model. (a) Inverse of the correlation length along the commensurate line. (b) Critical exponent $\nu$ as a function of the Ashkin-Teller asymmetry parameter $\lambda$. The dark blue line shows the exact result of Refs.\cite{kohmoto,obrien}. The open red circle is the numerical result of this work $\nu\simeq 0.777$ (result for $N=3001$ shown in panel (a)) and $\lambda\simeq 0.57$ as shown in Fig.\ref{fig:towers}. }
\label{fig:xi_comm}
\end{figure}

We extract the critical exponent correlation length along the commensurate line which, close to the transition, is given by $V_3/\Omega=0.3645\Delta/\Omega+2.825$.
 Since we expect a direct transition the critical exponent has to be the same on both sides of the critical point. However, the pre-factor is non-universal. We therefore fit our numerical data with:
$$|x-\Delta_c|^\nu\times [a \theta(x-\Delta_c)+b\theta(\Delta_c-x)],$$
where $a,b,\Delta_c$ and $\nu$ are fitting parameters; and $\theta(x)$ is the Heaviside function: $\theta(x)=1$ if $x>0$ and zero otherwise. The results are presented in Fig.\ref{fig:xi_comm}(a).

We compare the values of $\lambda$ and $\nu$ obtained to fit the hard-boson model with the conformal field theory result of Kohmoto et al.\cite{kohmoto,obrien} in Fig.\ref{fig:xi_comm}(b). The agreement is very good.

\textbf{Extraction of the critical exponent $\alpha$}.
In order to extract the specific heat critical exponent $\alpha$ we look at the divergence of the second derivative of the ground-state energy density $d^2e/d(\Delta/\Omega)^2$. In order to check the finite-size effects we take the energy per site $e=E/N$ extracted from the total ground-state energy $E$ for finite chains with two values of the number of sites: $N=1201$ and $N=2101$. We also consider the difference between the two ground-state energies $E_{2101}-E_{1201}$ to suppress  the edge effects and get a better estimate for the bulk energy per site as $(E_{2101}-E_{1201})/900$. Approaching the Ashkin-Teller point the specific heat should diverge as $|\Delta-\Delta_c|^\alpha$. The effective exponent $\alpha_\text{eff}$ close to the transition can thus be obtained as the slope of $\log d^2e/d(\Delta/\Omega)^2$ with respect to $\log |\Delta-\Delta_c|$.
The results are presented in Fig.\ref{fig:alpha}(a), where the pink line shows the location of the critical point. According to the hyperscaling relations $\alpha=2-2\nu$ and to our estimate of the correlation critical exponent $\nu\approx0.78\pm 0.02$, the specific heat critical exponent is expected to be $\alpha\approx0.44\pm 0.04$. This corresponds to the grey area in Fig.\ref{fig:alpha}(a), showing that our results for $\alpha$ are in reasonable agreement with this estimate at the Ashkin-Teller point. 

Far enough from the Ashkin-Teller point, the transition is expected to take place through an intermediate floating phase. At the Pokrovsky-Talapov point, the second derivative of the energy is expected to be very asymmetric, with a divergence with exponent $1/2$ on the incommensurate side and no divergence on the commensurate side \cite{HuseFisher1984}. The results of Fig.\ref{fig:alpha}(d) obtained for $V_3/\Omega=2$ are in good agreement with these predictions.

\begin{figure}[h!]
\centering 
\includegraphics[width=0.5\textwidth]{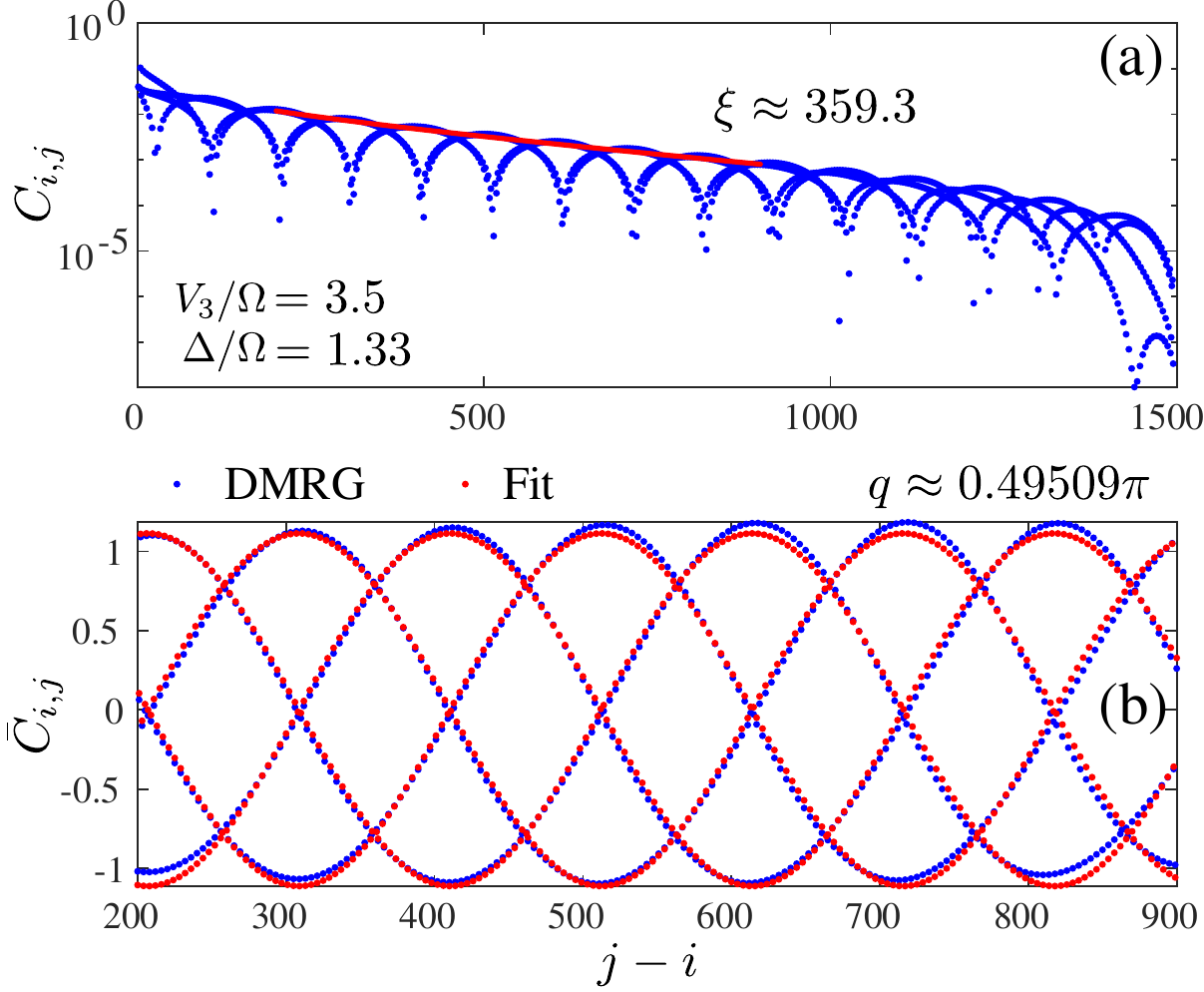}
\caption{ Example of  fit of the correlation function to the Ornstein-Zernicke form. In the first step (a), we extract the correlation length discarding the oscillations. In the second step (b), we fit the reduced correlation function to extract the wave-vector $q$.}
\label{fig:IncomFit}
\end{figure}

\textbf{Extraction of the correlation length and of the wave-vector.}
In order to extract the correlation length and the wave-vector $q$, we fit the boson-boson correlation function to the Ornstein-Zernicke form\cite{ornstein_zernike}:

\begin{equation}
C_\mathrm{OZ}{i,j}\propto \frac{e^{-|i-j|/\xi}}{\sqrt{|i-j|}}\cos(q|i-j|+\varphi_0),
\end{equation}
where the correlation length $\xi$, the wave vector $q$, and the initial phase $\varphi_0$ are fitting parameters. In order to extract the correlation length and the wave-vector with a sufficiently high precision, we fit the correlation function in two steps.
 First, we discard the oscillations and fit the main slope of the decay as shown in Fig.\ref{fig:IncomFit}. This allows us to perform a fit  in a semi-log scale $\log C(x=|i-j|)\approx c-x/\xi-\log(x)/2$, that in general provides more accurate estimates of the correlation length on a long scale. Second we define a reduced correlation function
 
 \begin{equation}
\tilde{C}_{i,j}=C_{i,j} \frac{\sqrt{|i-j|}}{e^{-|i-j|/\xi+c}}
\end{equation}
and fit it with a cosine $\tilde{C}_{i,j}\approx a\cos(q|i-j|+\varphi_0) $ as shown in Fig.\ref{fig:IncomFit}(b). The agreement is almost perfect: The DMRG data (blue dots) are almost completely behind the fit (red dots).

\section{Acknowledgments}
We thank Andreas L\"auchli and Samuel Nyckees for insigntful discussions.
This work has been supported by the Swiss National Science Foundation.
The calculations have been performed using the facilities of the University of Amsterdam.

\bibliographystyle{apsrev4-1}
\bibliography{bibliography,comments}

\newpage

\section{Supplementary Material for:\\ ``Kibble-Zurek exponent and chiral transition of the period-4 phase of Rydberg chains"}

\subsection{Blockade model and $1/R^6$ potential}

In this section we  further discuss the relevance of the $r$-site blockade model as an approximation of the Rydberg model with van der Waals interactions. This model is defined by the Hamiltonian in Eqs. (6-7) of the main text.

\begin{figure}[h!]
\centering 
\includegraphics[width=0.5\textwidth]{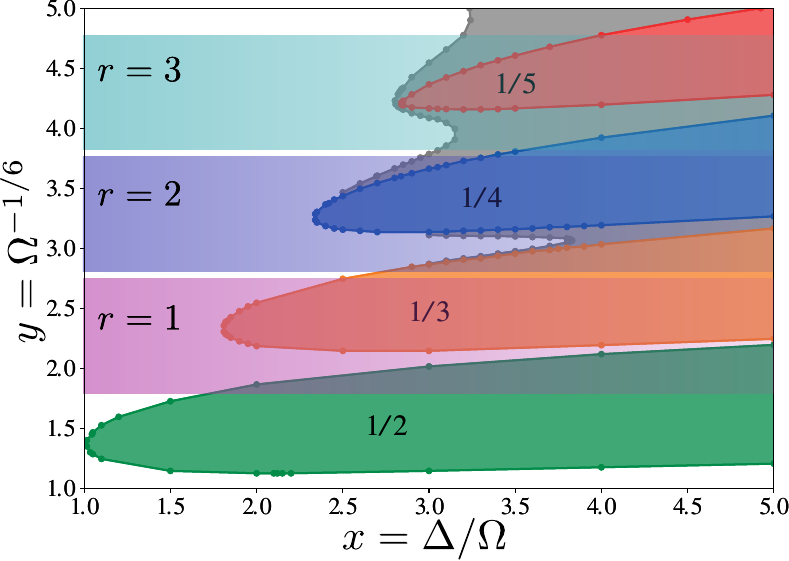}
\caption{Approximate regions where the blockade model is applicable. As a reference we use the phase diagram of the model with $1/R^6$ potential obtained in Ref.\onlinecite{rader2019floating}. In the used notations the coupling constant $V_1$ of the $1/R^6$ term is set to $V_1=1$.}
\label{fig:rader}
\end{figure}

In the absence of interaction between sites at distance $r+1$ and larger, the model with $r$-site blockade has two gapped phases - a disordered phase for positive or small negative values of the chemical potential $\Delta$ and a period $p=r+1$ crystalline phase that corresponds to the maximally occupied state consistent with the blockade. Upon increasing the repulsion  $V_{r+1}$ between sites at distance $r+1$, a period $p=r+2$ phase is expected to emerge. Thus the $r$-rite blockade model can effectively describe the transition to the $p=r+2$ phase and the region between the $p=r+1$ and $p=r+2$ phases. Note that in the blockade models, and unlike in  the Rydberg Hamiltonian of Eqs. (1-2) of the main text, we do not limit $V_{r+1}$ to be strictly positive.
In Fig.\ref{fig:rader} we marked the approximate regions where the blockade model is applicable. As a reference we use the phase diagram of the model with $1/R^6$ potential obtained in Ref.\onlinecite{rader2019floating}. 

\subsection{Further details on conformal towers at the Ashkin-Teller point}

\begin{figure}[t!]
\centering 
\includegraphics[width=0.4\textwidth]{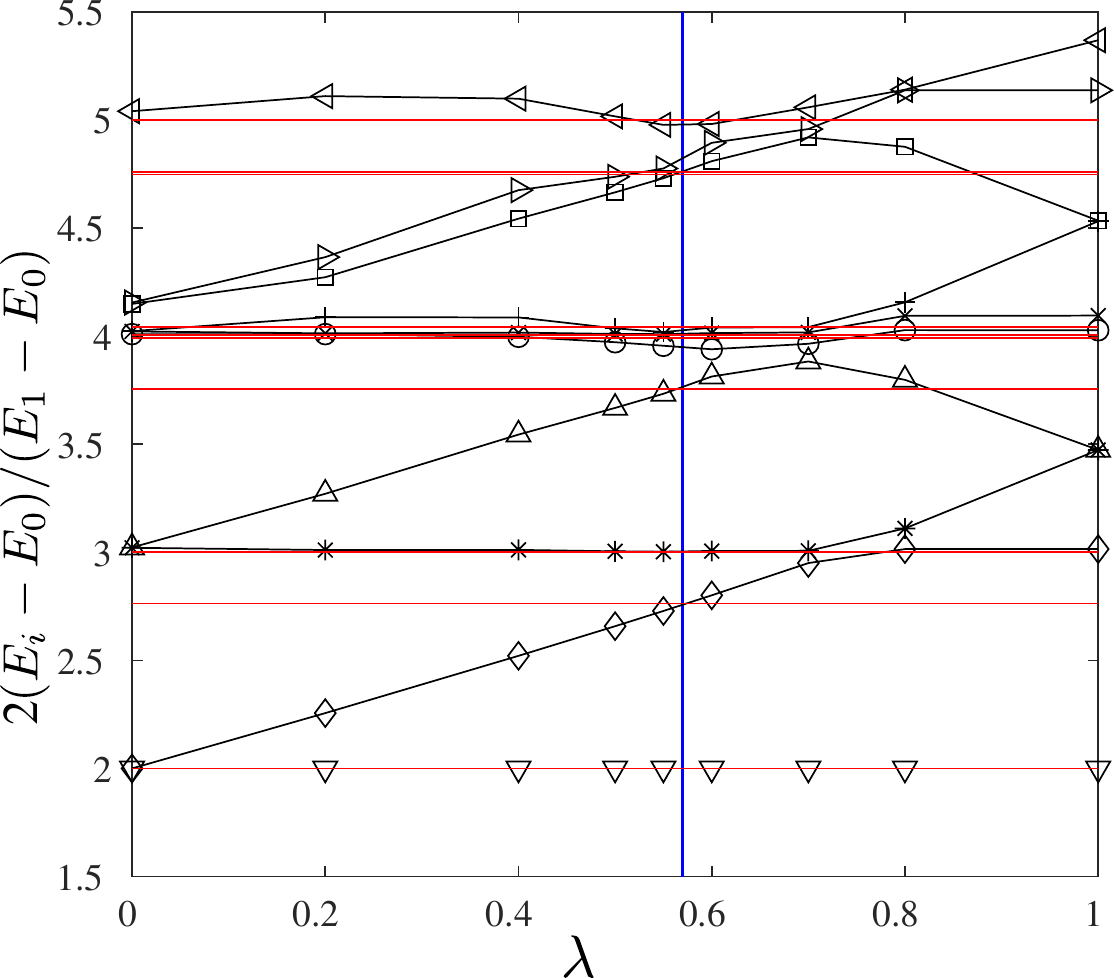}
\caption{Energy spectrum of the Ashkin-Teller model (black) as a function of $\lambda$ re-scaled with respect to its lowest state for a chain with $N=60$ sites and A-A boundary conditions. The results are compared to the re-scaled spectrum of the hard-boson model for $N=201$ sites with bosons on the first and last sites (red). We extract $\lambda$ by looking for the value where the spectra match best. Quite remarkably, the crossings of different levels occur for about the same value of $\lambda$, supporting the idea that the Ashkin-Teller model is the appropriate model to compare to. The resulting value of $\lambda$ corresponds to the blue line at $\lambda\approx0.57$.}
\label{fig:spectr}
\end{figure}

Here we show how the conformal towers of the blockade model and of the Ashkin-Teller model can be made systematic. In both the Ashkin-Teller and hard-boson models, the velocity of sound is a non-universal constant. So a quantitative comparison of these two spectra is only possible when the velocity is removed. This we achieve by re-scaling both spectra with respect to the lowest excited state, using a pre-factor 2 for a reason explained below. The crossing point of the two spectra gives the estimate of $\lambda\approx0.57$ as shown in Fig.\ref{fig:spectr}.

\begin{figure}[t!]
\centering 
\includegraphics[width=0.4\textwidth]{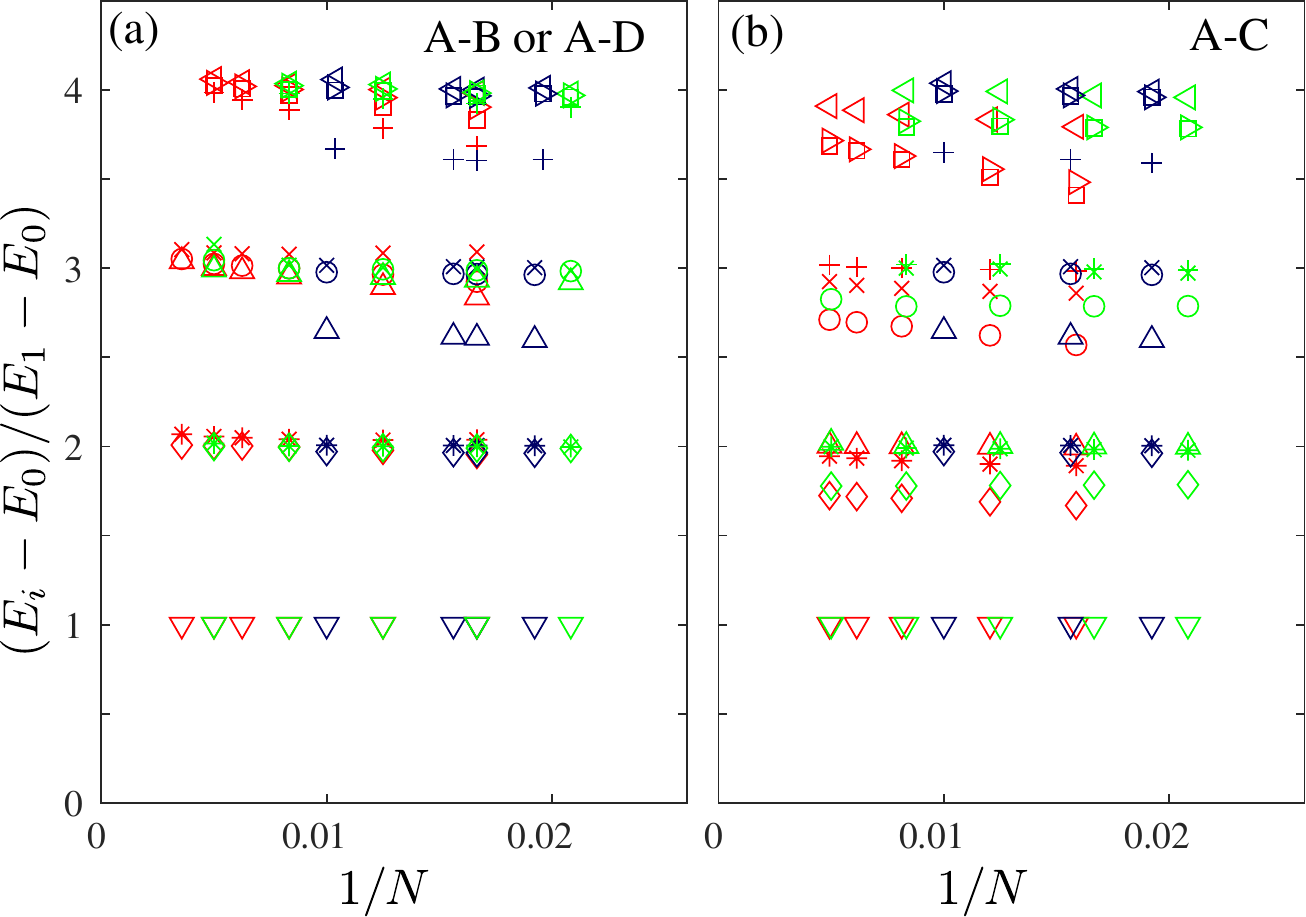}
\caption{ Conformal towers of states at the Ashkin-Teller point for A-B or A-D (a) and for A-C (b) boundary conditions. Red symbols state for the hard-boson data, green for the Ashkin-Teller model at $\lambda=0.57$, blue for the 4-state Potts model  (equivalent to  Ashkin-Teller at $\lambda=1$)}
\label{fig:ABADtowers}
\end{figure}

We further test the extracted value of $\lambda$ by looking at the finite-size scaling of the conformal towers of states. The results for A-A
 boundary conditions have been presented in the main text. In Fig.\ref{fig:ABADtowers} we present the results for A-B (same as A-D) and A-C boundary conditions. Note that here we do not use the pre-factor 2 for the re-scaled spectrum. This is related to the fact that for A-A boundary condition we expect the lowest energy states to be described by the identity conformal tower $I$ for which the $n=1$ excitation is missing. According to Fig.4(a) of the main text this seems to be the case everywhere for $0\leq \lambda \leq 1$. To the best of our knowledge the boundary-field correspondence has not been worked out in CFT for the Ashkin-Teller model, so the operator content for A-B, A-C and A-D boundary conditions is not known. However, by analogy with other minimal models, one can expect an equally spaced spectrum for the primary conformal tower, so the lowest state will be $n=1$. 
 
\subsection{Numerical data along a few selected cuts}

\begin{figure}[t!]
\centering 
\includegraphics[width=0.5\textwidth]{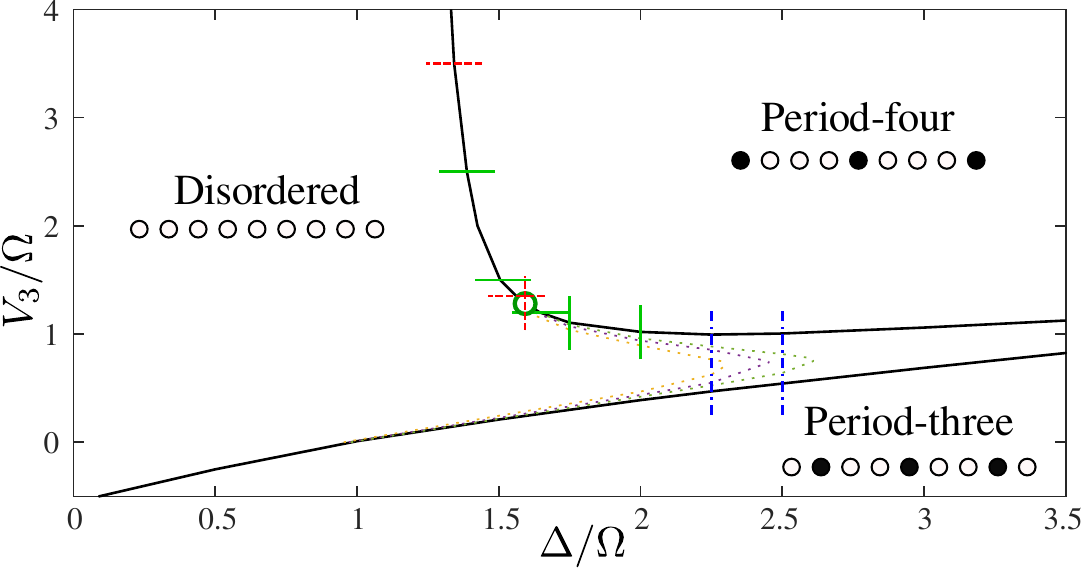}
\caption{Position of the cuts across the transition to the $p=4$ phase along which we have investigated the properties of the transition in detail. For clarity the length of each cut is enlarged. Solid black lines indicate the phase boundaries, the open green circle states for the symmetric Ashkin-Teller point. Dotted lines between $p=3$ and $p=4$ phases are equal-$\xi$ lines with $\xi=50$ (yellow), 100 (purple), 200 (green). Red dashed lines mark the cuts considered in the main text. Data across solid green lines are provided in Fig.\ref{fig:more_cuts}. Data along the cuts marked by blue dash-dotted lines are shown in Fig.\ref{fig:Um5}.}
\label{fig:phaseDiag_cuts}
\end{figure}

\begin{figure}[t!]
\centering 
\includegraphics[width=0.5\textwidth]{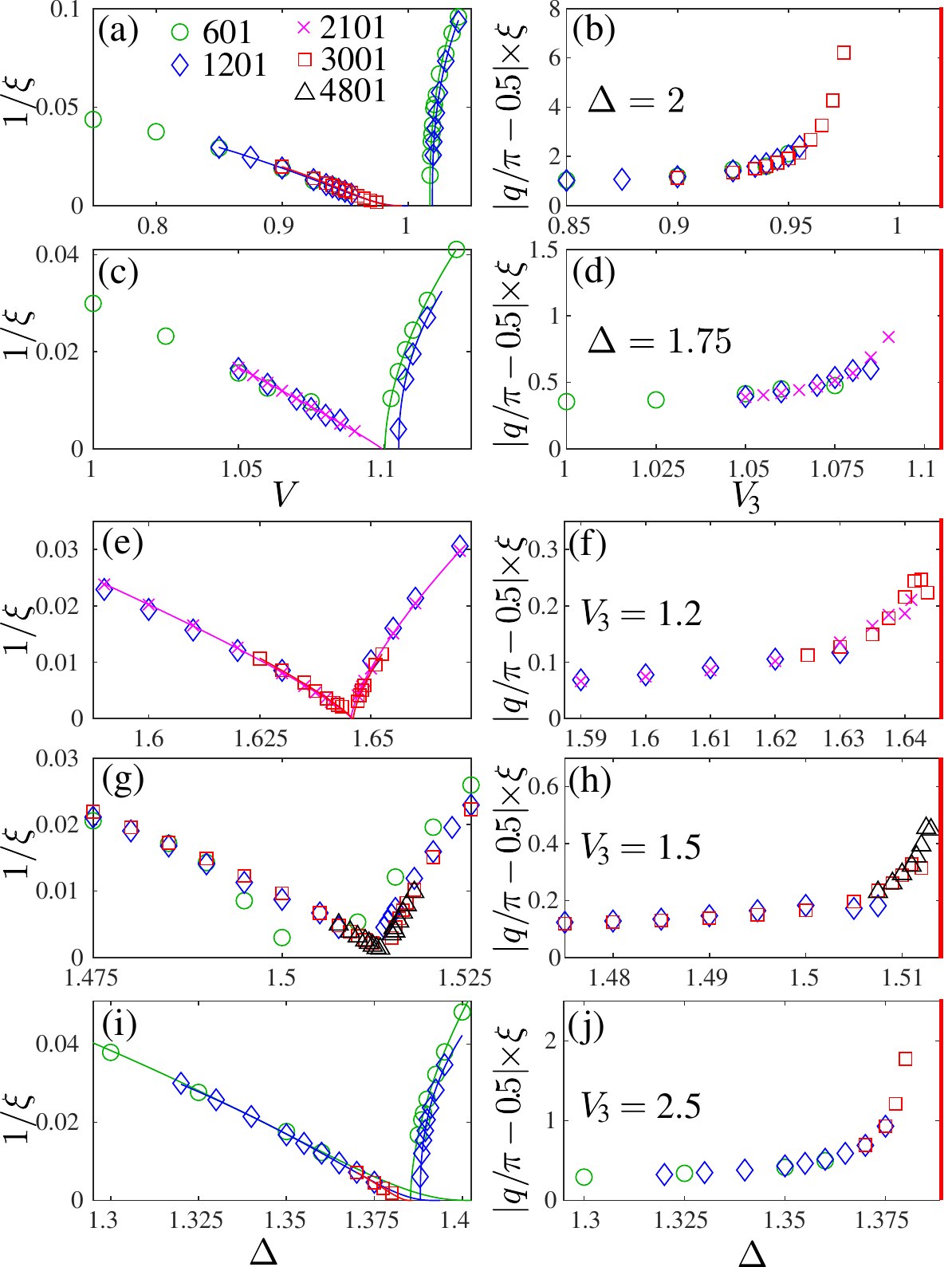}
\caption{Inverse of the correlation length (left) and product $|q\pi-1/2|\times\xi$ for a few selected horizontal (a-d) and vertical (e-j) cuts across the $p=4$ phase transition below (a-f) and above (g-j) the symmetric Ashkin-Teller point. Different symbols correspond to the system sizes listed in (a). }
\label{fig:more_cuts}
\end{figure}

In the main text we probe the nature of the phase transition to the $p=4$ phase based on numerical results for the correlation length $\xi$ and the wave-vector $q$ across three selected cuts. The location of these cuts is shown on the phase diagram in Fig.\ref{fig:phaseDiag_cuts} by red dashed lines. In Fig.\ref{fig:more_cuts} we provide data for five more cuts that cross the $p=4$ phase boundary at different places indicated in Fig.\ref{fig:phaseDiag_cuts} by green solid lines. 

The data for $\Delta/\Omega=2$ and for $V_3/\Omega=2.5$ point towards an intermediate floating phase. The data for $V_3/\Omega=1.2$ just below the Ashkin-Teller point is consistent with a chiral transition, similar to the data for $V_3/\Omega=1.35$ just above the Ashkin-Teller point (see main text). The data for $\Delta/\Omega=1.75$ and $V_3/\Omega=1.5$ are not conclusive: the product  $|q\pi-1/2|\times\xi$ increases by a factor of $\approx 2$ but seems to stay finite. If indeed there is a chiral transition very close to the Ashkin-Teller point, then these data might signal the proximity to the Lifshitz points at which a floating phase emerges.

\begin{figure}[t!]
\centering 
\includegraphics[width=0.5\textwidth]{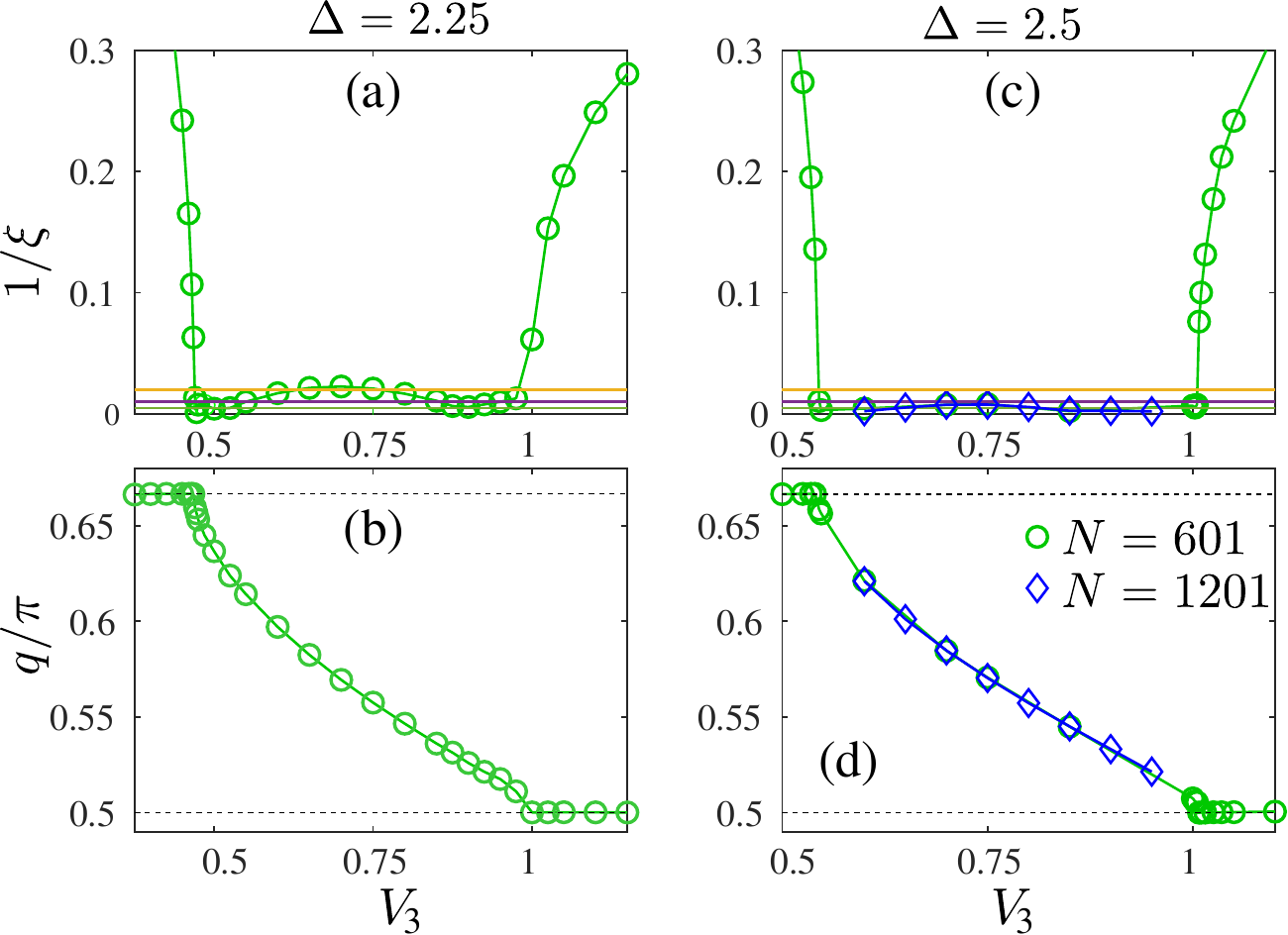}
\caption{Inverse of the correlation length (left) and wave-vector $q$ (right) across two vertical cuts from $p=3$ to $p=4$ phase. In (a) and (c) horizontal lines corresponding to $\xi=50$ (yellow), $\xi=100$ (purple) and $\xi=200$ (green) are shown for reference. }
\label{fig:Um5}
\end{figure}

In Fig.\ref{fig:Um5} we present the data across two different cuts that connect the $p=3$ and $p=4$ phases marked in Fig.\ref{fig:phaseDiag_cuts} by blue dash-dotted lines. We see that the correlation length remains very large between the two Pokrovsky-Talapov transitions: for $\Delta/\Omega=2.5$ it never goes below $\xi=120$. Beyond $\Delta/\Omega=2.5$ the finite-$\xi$ region cannot be resolved with our algorithm. 

Let us also comment on the transition out of the period-three phase. It is expected to take place either through  an intermediate floating phase or through a direct transition in the Huse-Fisher chiral universality class. Since the wave-vector $q>2\pi/3$ cannot be realized at finite $V_3$ because of the two-site blockade, the conformal 3-state Potts point is pushed to $V_3=-\infty$.

\end{document}